\def\year{2021}\relax
\documentclass[letterpaper]{article} 
\usepackage{aaai21}  
\usepackage{times}  
\usepackage{helvet} 
\usepackage{courier}  
\usepackage[hyphens]{url}  
\usepackage{graphicx} 
\usepackage{natbib}  
\usepackage{caption} 
\usepackage{cite}
\usepackage{amssymb} 
\usepackage{amstext}
\usepackage{subfigure}
\usepackage{booktabs}
\usepackage{enumitem}
\usepackage{multirow}
\usepackage[switch]{lineno}

\title{A Graph-based Relevance Matching Model for Ad-hoc Retrieval}
\author{\textbf{Yufeng Zhang\textsuperscript{\rm 1}\thanks{Equal contribution}, Jinghao Zhang\textsuperscript{\rm 1,\rm 2}\footnotemark[1], Zeyu Cui\textsuperscript{\rm 1,\rm 2}, Shu Wu\textsuperscript{\rm 1,\rm2,\rm 3}\thanks{Corresponding author} and Liang Wang\textsuperscript{\rm 1,\rm 2}} \\
}
\affiliations{\textsuperscript{\rm 1}Institute of Automation, Chinese Academy of Sciences \\
  \textsuperscript{\rm 2}University of Chinese Academy of Sciences \\
  \textsuperscript{\rm 3}Artificial Intelligence Research, Chinese Academy of Sciences \\
  \texttt{\{yufeng.zhang,jinghao.zhang\}@cripac.ia.ac.cn} \\ \texttt{\{zeyu.cui,shu.wu,wangliang\}@nlpr.ia.ac.cn}\\}

\begin{document}

\maketitle

\begin{abstract}
To retrieve more relevant, appropriate and useful documents given a query, finding clues about that query through the text is crucial. Recent deep learning models regard the task as a term-level matching problem, which seeks exact or similar query patterns in the document. However, we argue that they are inherently based on local interactions and do not generalise to ubiquitous, non-consecutive contextual relationships. In this work, we propose a novel relevance matching model based on graph neural networks to leverage the document-level word relationships for ad-hoc retrieval. In addition to the local interactions, we explicitly incorporate all contexts of a term through the graph-of-word text format. Matching patterns can be revealed accordingly to provide a more accurate relevance score. Our approach significantly outperforms strong baselines on two ad-hoc benchmarks. We also experimentally compare our model with BERT and show our advantages on long documents.

\end{abstract}

\section{Introduction}
Deep learning models have proved remarkably successful for information retrieval (IR) in recent years. The goal herein is to 
rank among a collection of documents the top relevant ones given a query. By utilising deep neural networks, these models aim to learn a function that can automatically extract matching patterns from two pieces of text, that is the query and the document, end-to-end in place of hand-crafted features. 

In general, there are two categories of neural matching architectures. One is called representation-based matching, which projects the query and document into the same low-dimensional semantic space and scores according to their similarity. Examples include DSSM \cite{huang2013learning}, ARC-I \cite{hu2014convolutional}, and CDSSM \cite{shen2014latent}. Another is called interaction-based matching, which learns relevant patterns directly from the interaction signals between the query and the document. Examples include DRMM \cite{guo2016deep}, KNRM \cite{xiong2017end}, 
and PACRR \cite{hui2017pacrr,hui2018co}. While the first category primarily concentrates on the semantics, the second emphasises more on the relevance. As discussed in \cite{guo2016deep}, there are significant differences between semantic matching and relevance matching. The latter is naturally more suitable for ad-hoc retrieval since the term-level query-document interaction provides more specific matching signals than the ensemble of semantic representations. 

\begin{figure}[t]
	\centering
	\subfigure[Query-document pair from Robust04]
	{\includegraphics[scale=0.4]{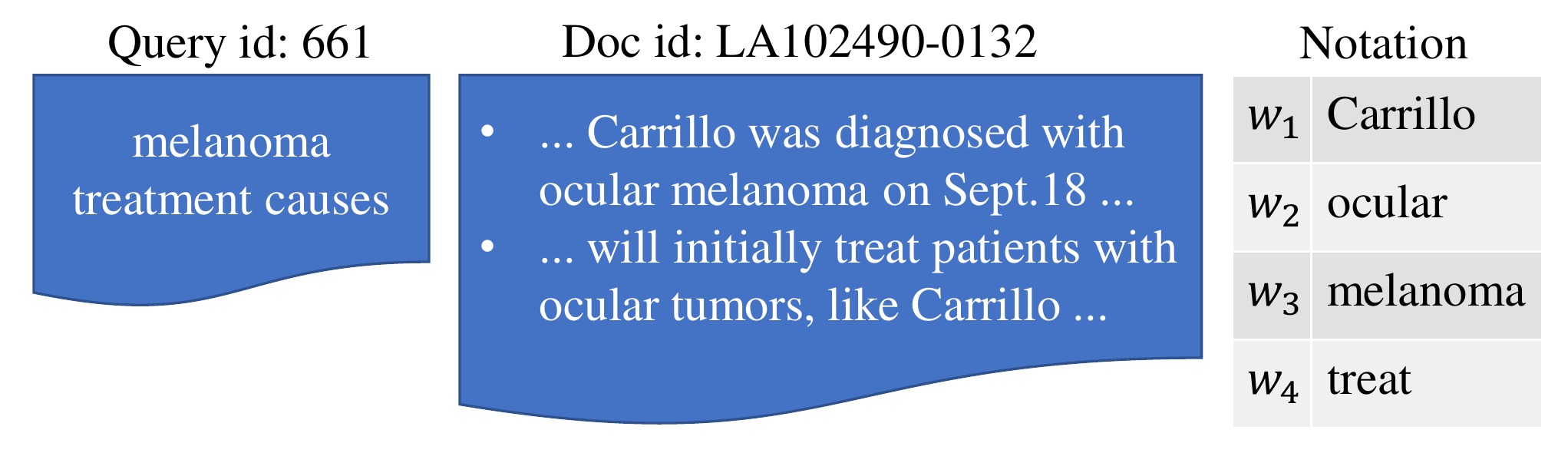}
	\label{fig:1a}}
	\subfigure[A local context scheme]
	{\includegraphics[scale=0.24]{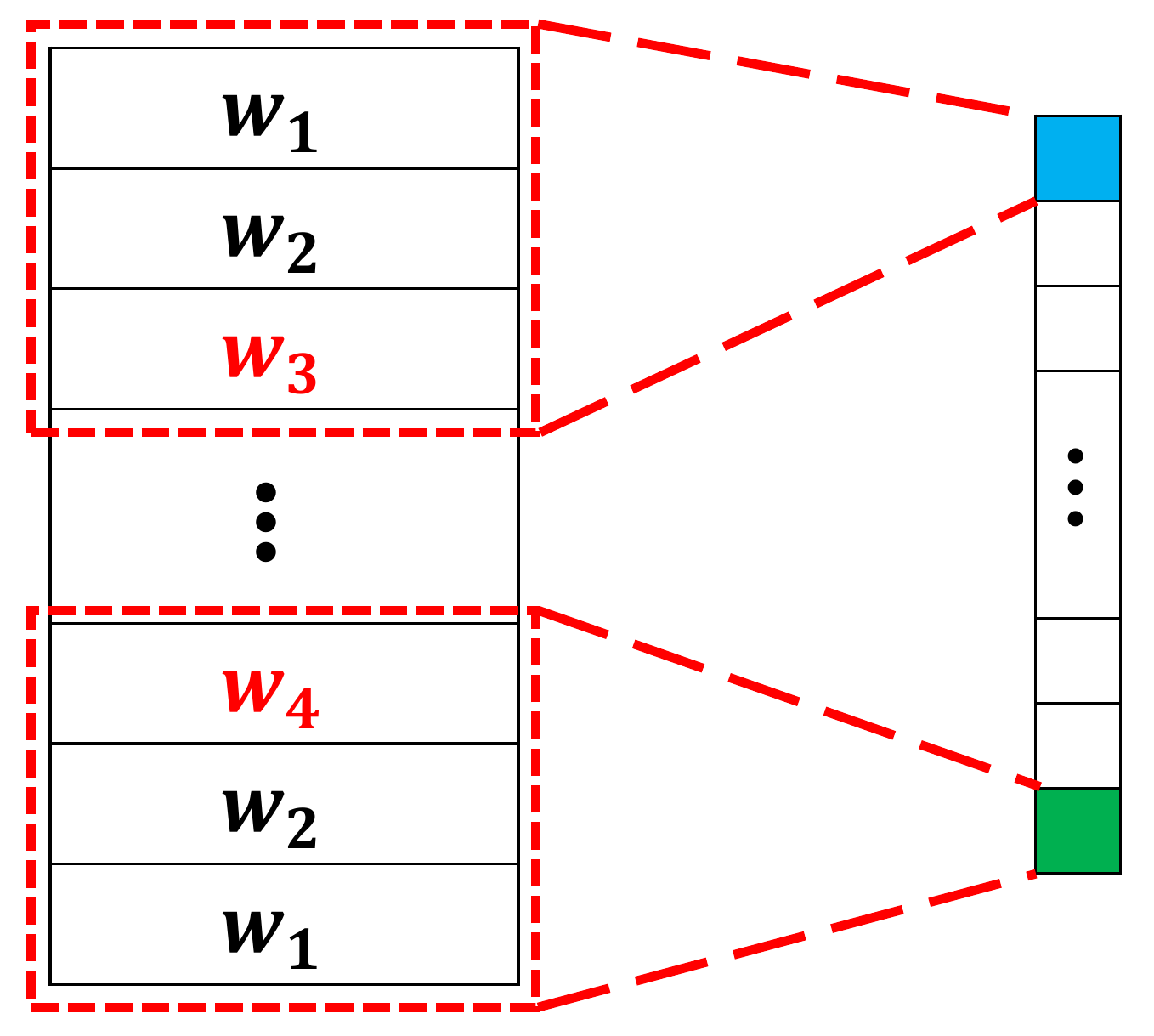}
	\label{fig:1b}}
	\subfigure[A graph-based context scheme]
	{\includegraphics[scale=0.22]{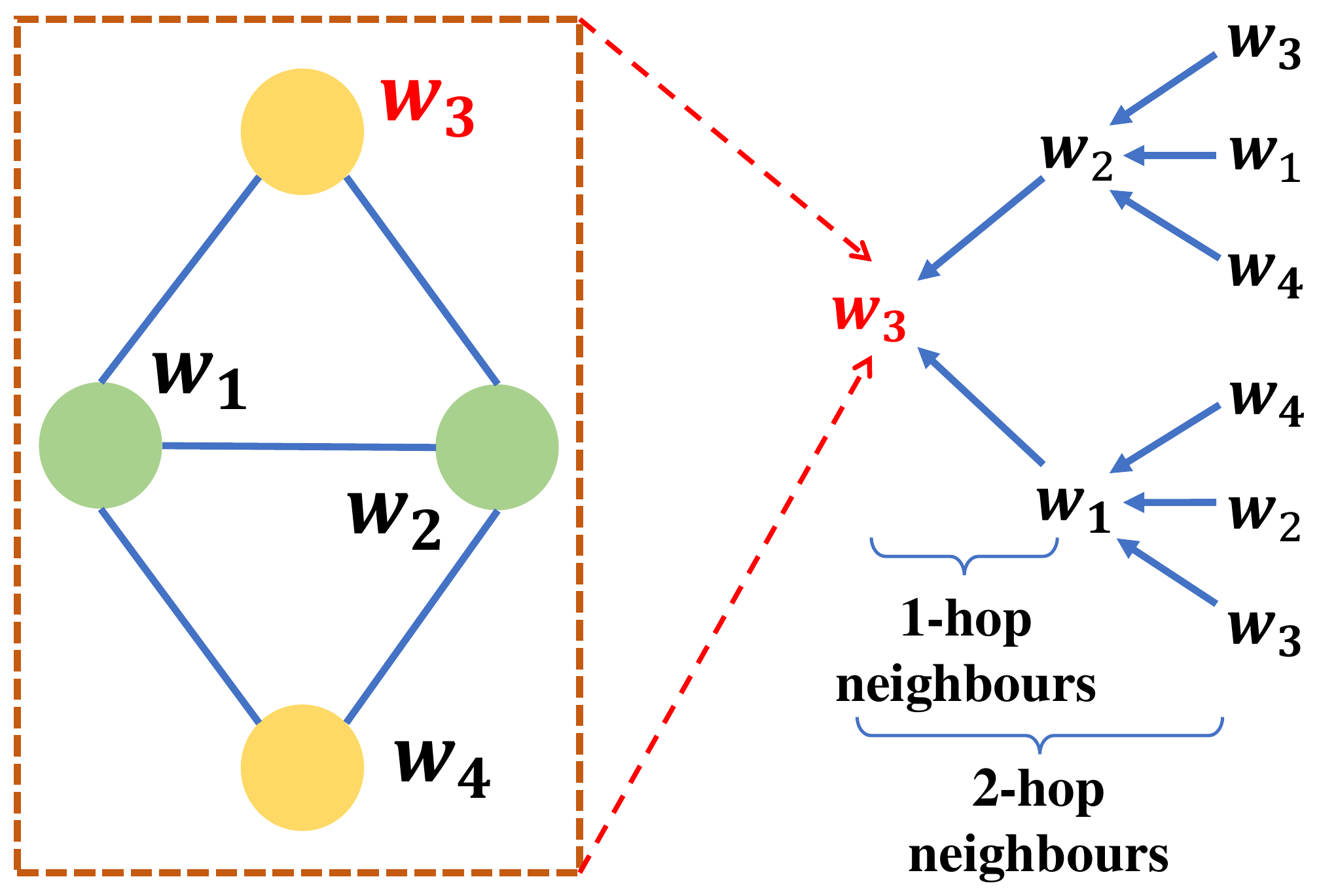}
	\label{fig:1c}}
	\label{fig:1} 
	\caption{An example of relevant query-document pair with two sentences far apart in the document (some words omitted). Local context scheme fails to discover the long-distance matching patterns due to the restriction of context. Graph-based context scheme works since words ``Carrillo'' and ``ocular'' play an important bridge role to connect ``melanoma" and ``treat" together.}
\end{figure}

In addition to the term-level query-document interaction, the document-level word relationships are also essential for relevance matching yet less explored so far. Taking Figure \ref{fig:1a} as an example, when searching with the query ``melanoma treatment", the retrieved document is expected to be highly relevant to them as a whole rather than to any single of ``melanoma" or ``treatment". However, query phrases do not always appear exactly in the document. It occurs more frequently that they (or their synonyms) distribute non-consecutively in any passage and still reserve a long-distance contextual association. Many works that rely on local word sequences \cite{pang2016text,pang2017deeprank,hui2017pacrr} fail to discover such dependencies due to the restriction of context, as illustrated in Figure \ref{fig:1b}. They, therefore, lead to a low score. We argue that these traditional term-level interactions are insufficient for relevance matching, and document-level relationships should be considered 
explicitly and concurrently. 

With recent researches towards graphs for natural language processing (NLP), \citet{yao2019graph} and \citet{zhang2020every} have demonstrated the usage of graph neural networks as a language model and their benefit in capturing long-distance word dependencies. Such graph structures could help search for non-consecutive phrases while maintaining their contextual meaning. For instance, Figure \ref{fig:1c} illustrates a connected graph for the document, where the words ``ocular" and ``Carrillo" nearby ``melanoma" and ``treat" could serve as a bridge connecting them. The query phrase emerges integrally in this way, resulting in a strong matching signal. Given the above, we aim to leverage the graph neural networks to expand the respective field through a flexible text format and assist in the document-level word relationships for ad-hoc retrieval.

In this work, we propose a \textbf{G}raph-based \textbf{R}elevance \textbf{M}atching \textbf{M}odel (GRMM) to resolve the match problem of long-distance terms. For a pair of query and document, we first transform the document into the graph-of-word form \cite{rousseau2015text}, where nodes are unique words, and edges are their co-occurrent linkages. Each node feature is assigned with the interaction between its word and query terms. Instead of raw word features, the interaction vector contains substantial matching signals, which is critical for relevance matching. We then apply graph neural networks to propagate these matching signals on the document graph. Thus the query-document interaction and intra-document word relationships can be modeled jointly. Finally, to estimate a relevance score, we adopt a $k$-max-pooling strategy for each query term to filter out irrelevant noisy information and feed their features into a dense neural layer.

We validate GRMM on two representative ad-hoc retrieval benchmarks, where empirical results show the effectiveness and rationality of GRMM. We also compare our model with BERT-based method, where we find that BERT potentially suffers from the same problem when the document becomes long. 

To sum up, the contributions of this work are as follows:
\begin{itemize}
	\item We point out the importance of explicitly considering long-distance word relationships for ad-hoc retrieval to enhance the query search.
	\item We propose a novel graph-based relevance matching model to address word relationships over the document, which can learn term-level and document-level matching signals jointly.
	\item We conduct comprehensive experiments to examine the effectiveness of GRMM and understand its working principle.
\end{itemize}

\section{Related Work}
In this section, we briefly review some existing neural matching models and graph neural networks.

\subsection{Neural Matching Models}
Most neural matching models fall within two categories: representation-focused models, e.g. DSSM \cite{huang2013learning}, ARC-I \cite{hu2014convolutional}, CDSSM \cite{shen2014latent}, and interaction-focused models, e.g. MatchPyramid \cite{pang2016text}, DRMM \cite{guo2016deep}, PACRR \cite{hui2017pacrr}, KNRM \cite{xiong2017end}.

The representation-focused models follow the representation learning approach adopted in many natural language processing tasks. Queries and documents are projected into the same semantic space individually. The cosine similarity is then used between their high-level text representations to produce the final relevance score. For example, DSSM \cite{huang2013learning}, one of the earliest neural relevance matching models, employs simple dense neural layers to learn high-level representations for queries and documents. To enhance the projecting function, ARC-I \cite{hu2014convolutional} and CDSSM \cite{shen2014latent} devoted much effort into convolutional layers later on. 

In comparison, interaction-focused methods model the two text sequences jointly, by directly exploiting detailed query-document interaction signals rather than high-level representations of individual texts. For example, DRMM \cite{guo2016deep} maps the local query-document interaction signals into a fixed-length histogram, and dense neural layers are followed to produce final ranking scores. \citet{xiong2017end} and \citet{dai2018convolutional} both use kernel pooling to extract multi-level soft match features. Many other works rely on convolutional layers or spatial GRU over interaction signals to extract ranking features  
such as \cite{pang2016text,pang2017deeprank,hui2017pacrr,hui2018co,fan2018modeling}, which considers just local word connections. 

There are also several studies investigating how to apply BERT in ranking, e.g.  \citet{dai2019deeper} and \citet{macavaney2019cedr}. A common approach is to concatenate the document and query text together and feed them into the next sentence prediction task, where the `[CLS]' token embeds the representation of the query-document pair. 
\begin{figure*}[h]
	\centering
	\includegraphics[width=\textwidth]{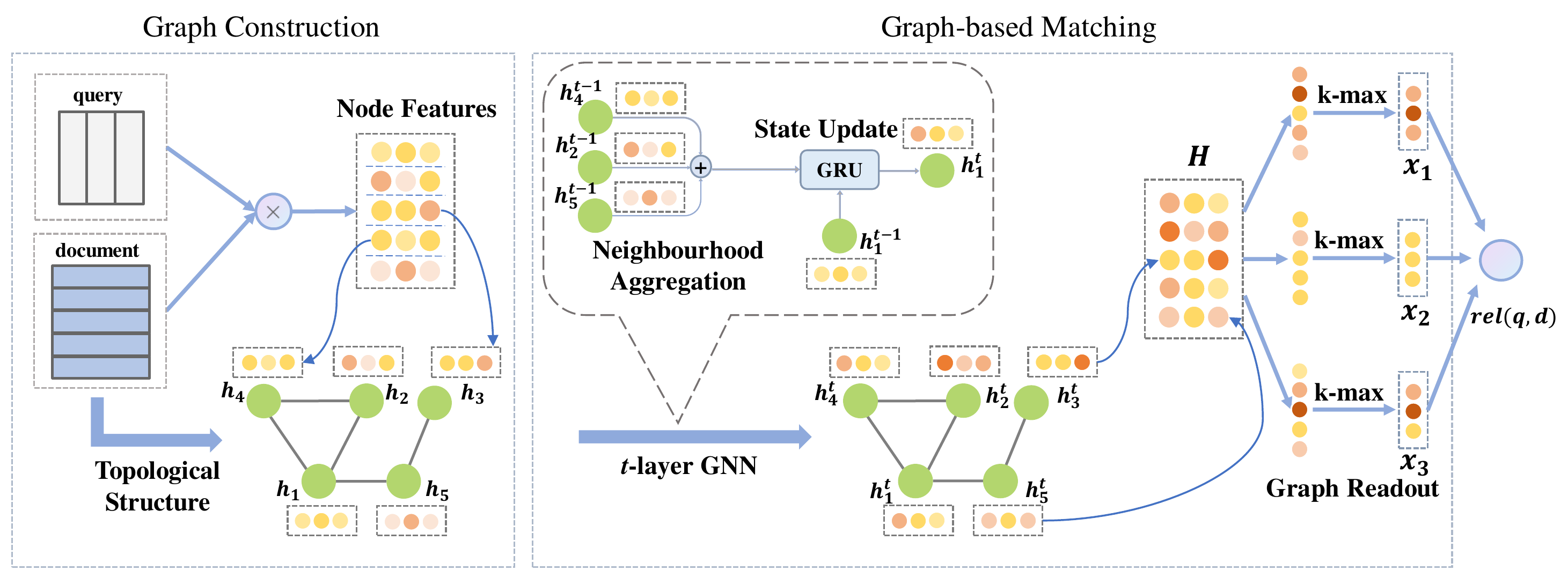}
	\caption{The workflow of the GRMM model. The document is first transformed into the graph-of-word form, where the node feature is the similarity between the word and each query term. Then, graph neural networks are applied to propagate these matching signals on the document graph. Finally, to estimate a relevance score, top-$k$ signals of each query term are chosen to filter out irrelevant noisy information, and their features are fed into a dense neural layer. }
	\label{fig:2} 
\end{figure*}

Nevertheless, the majority of existing neural matching models only take the linear text sequence, inevitably limiting the model capability. To this end, we propose to break the linear text format and represent the document in a flexible graph structure, where comprehensive interactions can be explicitly modeled.

\subsection{Graph Neural Networks}
Graph is a kind of data structure which cooperates with a set of objects (nodes) and their relationships (edges). Recently, researches of analysing graphs with machine learning have attracted much attention because of its great representative power in many fields. 

Graph neural networks (GNNs) are deep learning based methods that operate in the graph domain. The concept of GNNs is previously proposed by  \cite{scarselli2008graph}. Generally, nodes in GNNs update own hidden states by aggregating neighbourhood information and mixing things up into a new context-aware state. There are also many variants of GNNs with various kinds of aggregators and updaters, such as \cite{li2016gated,kipf2017semi,hamilton2017inductive,velivckovic2018graph}. 

Due to the convincing performance and high interpretability, GNNs have become a widely applied structural analysis tool. Recently, there are many applications covering from recommendation \cite{wu2019session,li2019fi} to NLP area, including text classification \cite{yao2019graph,zhang2020every}, question answering \cite{de2019question}, and spam review detection \cite{li2019spam}.

In this work, we employ GNNs in the relevance matching task to extract implicit matching patterns from the query-document interaction signals, which is intrinsically difficult to be revealed by existing methods.

\section{Proposed Method}

In this section, we introduce thoroughly our proposed Graph-based Relevance Matching Model (GRMM). We first formulate the problem and demonstrate how to construct the graph-of-word formation from the query and document, and then describe the graph-based matching method in details. Figure \ref{fig:2} illustrates the overall process of our proposed architecture.

\subsection{Problem Statement}
Given a query $q$ and a document $d$, they are represented as a sequence of  words 
$q=\left[w_{1}^{(q)}, \ldots, w_{M}^{(q)}\right]$  and $d=\left[w_{1}^{(d)}, \ldots, w_{N}^{(d)}\right]$, where $w_{i}^{(q)}$ denotes the $i$-th word in the query, $w_{i}^{(d)}$ denotes the $i$-th word in the document, $M$ and $N$ denote the length of the query and the document respectively.
The aim is to compute a relevance score $rel(q,d)$ regarding the query words and the document words.

\subsection{Graph Construction}
\label{sec:graphconstruct}
To leverage the long-distance term dependency information, the first step is to construct a graph $\mathcal{G}$ for the document. It typically consists of two components denoted as $\mathcal{G}=(\mathcal{V}, \mathcal{E})$, 
where $\mathcal{V}$ is the set of vertexes with \emph{node features}, and $\mathcal{E}$ is the set of edges as the \emph{topological structure}.

\subsubsection{Node features.}
We represent each unique word instead of sentence or paragraph in the document as a node. Thus the word sequence is squeezed to a node set $\left\{w_{1}^{(d)}, \ldots, w_{n}^{(d)}\right\}$, where $n$ is the number of unique words in the document ($|\mathcal{V}| = n  \leq N$). Each node feature is set the interaction signal between its word embedding and query term embeddings. We simply employ the cosine similarity matrix as the interaction matrix, denoted as $\mathbf{S} \in \mathbb{R}^{n \times M}$, where each element $\mathbf{S}_{ij}$ between document node $w^{(d)}_i$ and query term $w^{(q)}_j$ is defined as:
\begin{equation}\mathbf{S}_{i j}=cosine\left(\mathbf{e}_i^{(d)}, \mathbf{e}_j^{(q)}\right)
\end{equation}
where $\mathbf{e}_{i}^{(d)}$ and $\mathbf{e}_{j}^{(q)} $ are embedding vectors for $w_{i}^{(d)}$ and $w_{j}^{(q)}$ respectively. In this work, we use word2vec \cite{mikolov2013distributed} technique to convert words into dense and semantic embedding vectors.

\subsubsection{Topological structure.}
In addition to the node feature matrix, the adjacency matrix representing the topological structure constitutes for the graph as well. The structure generally describes the connection between the nodes and reveals their relationships. We build bi-directional connections for each pair of word nodes that co-occur within a sliding window, along with the original document word sequence $d$. By restricting the size of the window, every word can connect with their neighbourhood words which may share related contextual meanings. However, GRMM differs from those local relevance matching methods in that the combined word node can bridge all neighbourhoods together and therefore possess a document-level receptive field. In other words, it breaks the constraints of local context and can model the long-distance word dependencies that we concern. Note that in the worst case where there are no duplicate words, the graph would still perform as a sequential and local scheme. 

Formally, the adjacency matrix $\mathbf{A} \in \mathbb{R}^{n \times n}$ is defined as:
\begin{equation}
\mathbf{A}_{i j}=\left\{\begin{array}{ll}
count(i, j) & \text{if } i \not= j \\
0 & \text{otherwise}
\end{array}\right.
\end{equation}
where $count(i, j)$ is the number of times that the words $w_{i}^{(d)}$ and $w_{j}^{(d)}$ appear in the same sliding window. To alleviate the exploding/vanishing gradient problem \cite{kipf2017semi}, we normalise the adjacency matrix as $\tilde{\mathbf{A}} = \mathbf{D}^{-\frac{1}{2}} \mathbf{A} \mathbf{D}^{-\frac{1}{2}}$, where $\mathbf{D} \in \mathbb{R}^{n \times n}$ is the diagonal degree matrix and $\mathbf{D}_{ii} = \sum_j \mathbf{A}_{ij}$.

\subsection{Graph-based Matching}
Once we obtain the graph $\mathcal{G}$, we focus on making use of its node features and structure information with graph neural networks. In particular, the query-document interaction and the intra-document word interaction are learned mutually following the procedures - \emph{neighbourhood aggregation}, \emph{state update} and \emph{feature election}. 

\subsubsection{Neighbourhood Aggregation.}
As discussed in Section \ref{sec:graphconstruct}, we initialise the node state $\mathbf{h}^0_i$ with the query-document interaction matrix:
\begin{equation}\mathbf{h}^0_i =  \mathbf{S}_{i,:}
\end{equation}
where $\forall i\in [1, n]$ denotes the $i$-th node in the graph, and $\mathbf{S}_{i,:}$ is the $i$-th row of the interaction matrix $\mathbf{S}$.

Assume each word node either holds the core information or serves as a bridge connecting others, it is necessary to make the information flow and enrich the related fractions on the graph.
Through propagating the state representations to a node from its neighbours, it can receive the contextual information within the first-order connectivity as:
\begin{equation}\mathbf{a}_{i}^{t}=\sum_{(w_{i}, w_{j}) \in \mathcal{E}} \mathbf{\tilde{A}}_{ij} \mathbf{W}_{a} \mathbf{h}_{j}^{t}\end{equation}
where $\mathbf{a}_i^t \in \mathbb{R}^{M}$ denotes the summed message from neighbours, $t$ denotes the current timestamp, and $\mathbf{W}_a$ is a trainable transformation matrix to project features into a new relation space. When aggregate $t$ times recursively, a node can receive the information propagated from its $t$-hop neighbours. In this way, the model can achieve \emph{high-order aggregation} of the query-document interaction as well as the intra-document interaction.

\subsubsection{State Update.}
To incorporate the contextual information into the word nodes, we engage a GRU-like function \cite{li2016gated} to automatically adjust the merge proportion of its current representation $\mathbf{h}^{t}_i$ and the received representation $\mathbf{a}^{t}_i$, which is formulated as:
\begin{equation}\begin{array}{l}
\mathbf{z}_{i}^{t}=\sigma\left(\mathbf{W}_{z} \mathbf{a}_{i}^{t}+\mathbf{U}_{z} \mathbf{h}_{i}^{t}+\mathbf{b}_{z}\right)
\end{array}\end{equation}
\begin{equation}
\mathbf{r}_{i}^{t}=\sigma\left(\mathbf{W}_{r} \mathbf{a}_{i}^{t}+\mathbf{U}_{r} \mathbf{h}_{i}^{t}+\mathbf{b}_{r}\right)
\end{equation}
\begin{equation}\tilde{\mathbf{h}}_{i}^{t}=\tanh \left(\mathbf{W}_{h} \mathbf{a}_{i}^{t}+\mathbf{U}_{h}\left(\mathbf{r}_{i}^{t} \odot \mathbf{h}_{i}^{t}\right)+\mathbf{b}_{h}\right)\end{equation}
\begin{equation}\mathbf{h}_{i}^{t+1}=\tilde{\mathbf{h}}_{i}^{t} \odot \mathbf{z}_{i}^{t}+\mathbf{h}_{i}^{t} \odot\left(1-\mathbf{z}_{i}^{t}\right)\end{equation}
where $\sigma(\cdot)$ is the sigmoid function, $\odot$ is the Hardamard product operation, tanh$(\cdot)$ is the non-linear tangent hyperbolic activation function, and all $\mathbf{W_*, U_*}$ and $\mathbf{b_*}$ are trainable weights and biases. 

Specifically, $\mathbf{r}^{t}_i$ determines irrelevant information for hidden state $\tilde{\mathbf{h}}^{t}_i$ to forget (reset gate), while $\mathbf{z}^{t}_i$ determines which part of past information to discard and which to push forward (update gate). With the layer $t$ going deep, high-order information becomes complicated, and it is necessary to identify useful dependencies with the two gates. We have also tried plain updater such as GCN \cite{kipf2017semi} in our experiments but did not observe satisfying performance due to its simplicity.

\subsubsection{Graph Readout.}
The last phase involves locating the position where relevance matching happens as a delegate for the entire graph. Since it is suggested that not all words make contributions, and some may cause adverse influences \cite{guo2016deep}, here we only select the most informative features to represent the query-document matching signals. Intuitively, higher similarity means higher relevance possibility. Hence we perform a $k$-max-pooling strategy over the query dimension and select the top $k$ signals for each query term, which also prevents the model from being biased by the document length. The formulas are expressed as:
\begin{equation}\mathbf{H}=\mathbf{h}_{1}^{t} \parallel \mathbf{h}_{2}^{t} \parallel \ldots \parallel \mathbf{h}_{n}^{t}\end{equation}
\begin{equation}
\mathbf{x}_{j} = {topk}(\mathbf{H}_{:,j})
\end{equation}
where $\forall j\in [1, M]$ denotes the $j$-th query term, and $\mathbf{H}_{:,j}$ is the $j$-th column of the feature matrix $\mathbf{H}$.

\subsection{Matching Score and Training}
After obtaining low-dimensional and informative matching features $\mathbf{x}_j$, we move towards converting them into actual relevance scores for training and inference. Considering different terms may have different importances \cite{guo2016deep}, we assign each with a soft gating network as:
\begin{equation}g_{j}=\frac{\exp \left({c} \cdot idf_j \right)}{\sum_{j=1}^{M} \exp \left({c} \cdot idf_j \right)}\end{equation}
where $g_j$ denotes the term weight, $idf_j$ is the inverse document frequency of the $j$-th query term, and $c$ is a trainable parameter. To reduce the amount of parameters and avoid over-fitting, we score each query term with a weight-shared multi-layer perceptron (MLP) and sum them up as the final result:
\begin{equation}{rel}(q, d)=\sum_{j=1}^{M} g_j \cdot \tanh \left(\mathbf{W}_x \mathbf{x}_{j}+{b}_x \right)\end{equation}
where $\mathbf{W}_x, b_x$ are trainable parameters for MLP.

Finally, we adopt the pairwise hinge loss which is commonly used in information retrieval to optimise the model parameters:
\begin{small}
	\begin{equation}\mathcal{L}\left(q, d^{+}, d^{-}\right)=\max \left(0, 1-rel\left(q, d^{+}\right)+rel\left(q, d^{-}\right)\right)\end{equation} 
\end{small}
where $\mathcal{L}\left(q, d^{+}, d^{-}\right)$ denotes the pairwise loss based on a triplet of the query $q$, a relevant (positive) document sample $d^+$, and an irrelevant (negative) document sample $d^-$.

\section{EXPERIMENTS}
In this section, we conduct experiments on two widely used datasets to answer the following research questions:
\begin{itemize}
	\item RQ1: How does GRMM perform compared with different retrieval methods (typically traditional, local interaction-based, and BERT-based matching methods)?
	\item RQ2: How effective is the graph structure as well as the long-dependency in ad-hoc retrieval?
	\item RQ3: How sensitive (or robust) is GRMM with different hyper-parameter settings?
\end{itemize}

\subsection{Experiment Setup}
\subsubsection{Datasets.}
We evaluate our proposed model on two datasets: Robust04 and ClueWeb09-B.
\begin{itemize}
    \item Robust04\footnote{https://trec.nist.gov/data/cd45/index.html} is a standard ad-hoc retrieval dataset with 0.47M documents and 250 queries, using TREC disks 4 and 5 as document collections.
    \item ClueWeb09-B\footnote{https://lemurproject.org/clueweb09/} is the "Category B" subset of the full web collection ClueWeb09. It has 50M web pages and 200 queries, whose topics are accumulated from TREC Web Tracks 2009-2012.
\end{itemize}
Table \ref{tab:1} summarises the statistic of the two collections. For both datasets, there are two available versions of the query: a keyword title and a natural language description. In our experiments, we only use the title for each query.

\subsubsection{Baselines.}
To examine the performance of GRMM, we take three categories of retrieval models as baselines, including traditional (QL and BM25), local interaction-based (MP, DRMM, KNRM, and PACRR), and BERT-based (BERT-MaxP) matching methods, as follows: 

\begin{itemize}
    \item \textbf{QL} (Query likelihood model) \cite{zhai2004study} is one of the best performing language models that based on Dirichlet smoothing.
    \item \textbf{BM25} \cite{robertson1994some} is another effective and commonly used classical probabilistic retrieval model.
    \item \textbf{MP} (MatchPyramid) \cite{pang2016text} employs CNN to extract the matching features from interaction matrix, and dense neural layers are followed to produce final ranking scores.
    \item \textbf{DRMM} \cite{guo2016deep} performs a histogram pooling over the local query-document interaction signals. 
    \item \textbf{KNRM} \cite{xiong2017end} introduces a new kernel-pooling technique that extracts multi-level soft matching features.
    \item \textbf{PACRR} \cite{hui2017pacrr} uses well-designed convolutional layers and $k$-max-pooling layers over the interaction signals to model sequential word relations in the document.
    \item \textbf{Co-PACRR} \cite{hui2018co} is a context-aware variant of PACRR that takes the local and global context of matching signals into account.
    \item \textbf{BERT-MaxP} \cite{dai2019deeper} applies BERT to provide deeper text understanding for retrieval. The neural ranker predicts the relevance for each passage independently, and the document score is set as the best score among all passages.
\end{itemize}

\begin{table}[]
	\footnotesize
	\begin{tabular}{@{}ccccc@{}}
		\toprule
		\textbf{Dataset}     & \textbf{Genre} & \textbf{\# of qrys} & \textbf{\# of docs} & \textbf{avg.length} \\ \midrule
		\textbf{Robust04}    & news           & 250                 & 0.47M                & 460                         \\
		\textbf{ClueWeb09-B} & webpages       & 200                 & 50M                 & 1506                        \\ \bottomrule
	\end{tabular}
	\caption{Statistics of datasets.}
	\label{tab:1}
\end{table}

\subsubsection{Implementation Details.}
All document and query words were white-space tokenised, lowercased, and lemmatised using the WordNet\footnote{https://www.nltk.org/howto/wordnet.html}. We discarded stopwords as well as low-frequency words with less than ten occurrences in the corpus. Regarding the word embeddings, we trained 300-dimensional vectors with the Continuous Bag-of-Words (CBOW) model \cite{mikolov2013distributed} on Robust04 and ClueWeb-09-B collections. For a fair comparison, the other baseline models shared the same embeddings, except those who do not need. Implementation of baselines followed their original paper.

Both datasets were divided into five folds. We used them to conduct 5-fold cross-validation, where four of them are for tuning parameters, and one for testing \cite{macavaney2019cedr}. The process repeated five times with different random seeds each turn, and we took an average as the performance.

We implemented our method in PyTorch\footnote{Our code is at https://github.com/CRIPAC-DIG/GRMM}. The optimal hyper-parameters were determined via grid search on the validation set: the number of graph layers $t$ was searched in \{1, 2, 3, 4\}, the $k$ value of $k$-max-pooling was tuned in \{10, 20, 30, 40, 50, 60, 70\}, the sliding window size in \{3,5,7,9\}, the learning rate in \{0.0001, 0.0005, 0.001, 0.005, 0.01\}, and the batch size in \{8, 16, 32, 48, 64\}.
Unless otherwise specified, we set $t$ = 2 and $k$ = 40 to report the performance (see Section \ref{sec:neighbouraggre} and \ref{sec:featureelect} for different settings), and the model was trained with a window size of 5, a learning rate of 0.001 by Adam optimiser for 300 epochs, each with 32 batches times 16 triplets. All experiments were conducted on a Linux server equipped with 8 NVIDIA Titan X GPUs.

\subsubsection{Evaluation Methodology.}
Like many ad-hoc retrieval works, we adopted a re-ranking strategy that is more efficient and practical than ranking all query-document pairs. In particular, we re-ranked top 100 candidate documents for each query that were initially ranked by BM25. To evaluate the re-ranking result, we used the normalised discounted cumulative gain at rank 20 (nDCG@20) and the precision at rank 20 (P@20) as evaluation matrices.

\subsection{Model Comparison (RQ1)}
Table \ref{tab:2} lists the overall performance of different models, from which we have the following observations:
\begin{itemize}
	\item GRMM significantly outperforms traditional and local interaction-based models, and it is comparable to BERT-MaxP, though without massive external pre-training. To be specific, GRMM advances the performance of nDCG@20 by 14.4\% on ClueWeb09-B much more than by 5.4\% on Robust04, compared to the best-performed baselines excluding BERT-MaxP. It is reasonably due to the diversity between the two datasets. ClueWeb09-B contains webpages that are usually long and casual, whereas Robust04 contains news that is correspondingly shorter and formal. It suggests that useful information may have distributed non-consecutively, and it is beneficial to capture them together, especially for long documents. GRMM can achieve long-distance relevance matching through the graph structure regardless of the document length. 
	
	\item On the contrary, BERT-MaxP performs relatively better on Robust04 than on ClueWeb09-B. We explain the observation with the following two points. First, since the input sequence length is restricted by a maximum of 512 tokens, BERT has to truncate those long documents from ClueWeb09-B into several passages. It, therefore, loses relations among different passages, i.e. the long-distance dependency. Second, documents from Robust04 are generally written in formal languages. BERT primarily depends on the pre-trained semantics, which could naturally gain benefit from that. 
	
	\item Regarding the local interaction-based models, their performances slightly fluctuate around the initial ranking result by BM25. However, exceptions are DRMM and KNRM on ClueWeb09-B, where the global histogram and kernel pooling strategy may cause the difference. It implies that the local interaction is insufficient in ad-hoc retrieval task. Document-level information also needs to be considered. 
	
	\item Traditional approaches like QL and BM25 remain a strong baseline though quite straightforward, which means the exact matching of terms is still of necessity as \citet{guo2016deep} proposed. These models also avoid the problem of over-fitting, since they do not require parameter optimisation. 
\end{itemize}                       

\label{sec:modelcompare}
\begin{table}[]
	\fontsize{9.3pt}{11pt}\selectfont
    \begin{tabular}{@{}cllll@{}}
    \toprule
    \multirow{2}{*}{Model} & \multicolumn{2}{c}{Robust04}                           & \multicolumn{2}{c}{ClueWeb09-B}                        \\ \cmidrule(l){2-5} 
                           & \multicolumn{1}{c}{nDCG@20} & \multicolumn{1}{c}{P@20} & \multicolumn{1}{c}{nDCG@20} & \multicolumn{1}{c}{P@20} \\ \midrule
    QL                     & 0.415$^-$                   & 0.369$^-$                & 0.224$^-$                   & 0.328$^-$                \\
    BM25                   & 0.418$^-$                   & 0.370$^-$                & 0.225$^-$                   & 0.326$^-$                \\ \midrule
    MP                     & 0.318$^-$                   & 0.278$^-$                & 0.227$^-$                   & 0.262$^-$                \\
    DRMM                   & 0.406$^-$                   & 0.350$^-$                & 0.271$^-$                   & 0.324$^-$                \\
    KNRM                   & 0.415$^-$                   & 0.359$^-$                & 0.270$^-$                   & 0.330$^-$                \\
    PACRR                  & 0.415$^-$                   & 0.371$^-$                & 0.245$^-$                   & 0.278$^-$                \\
    Co-PACRR               & 0.426$^-$                   & 0.378$^-$                & 0.252$^-$                   & 0.289$^-$                \\ \midrule
    BERT-MaxP              & \textbf{0.469}                       & -                        & 0.293                       & -                        \\ \midrule
    GRMM                   & 0.449                        & \textbf{0.387}                    & \textbf{0.310}                       & \textbf{0.354}                    \\ \bottomrule
    \end{tabular}
	\caption{Performance comparison of different methods. The best performances on each dataset and metric are highlighted. Significant performance degradation with respect to GRMM is indicated (-) with p-value $\leq$ 0.05.}
	\label{tab:2}
\end{table}

\subsection{Study of Graph Structure (RQ2)}
\label{sec:graphstructure}
To dig in the effectiveness of the document-level word relationships of GRMM, we conduct further ablation experiments to study their impact. Specifically, we keep all settings fixed except substituting the adjacency matrix with: 
\begin{itemize}
	\item \textbf{Zero matrix}: Word nodes can only see themselves, and no neighbourhood information is aggregated. This alternative can be viewed as not using any contextual information. The model learns directly from the query-document term similarity.
	\item \textbf{Word sequence}, the original document format: No words are bound together, and they can see themselves as well as their previous and next ones. This alternative can be viewed as only using local contextual information. It does not consider long-distance dependencies. 
\end{itemize}

\begin{figure}[h]
	\centering
	\includegraphics[width=.47\textwidth]{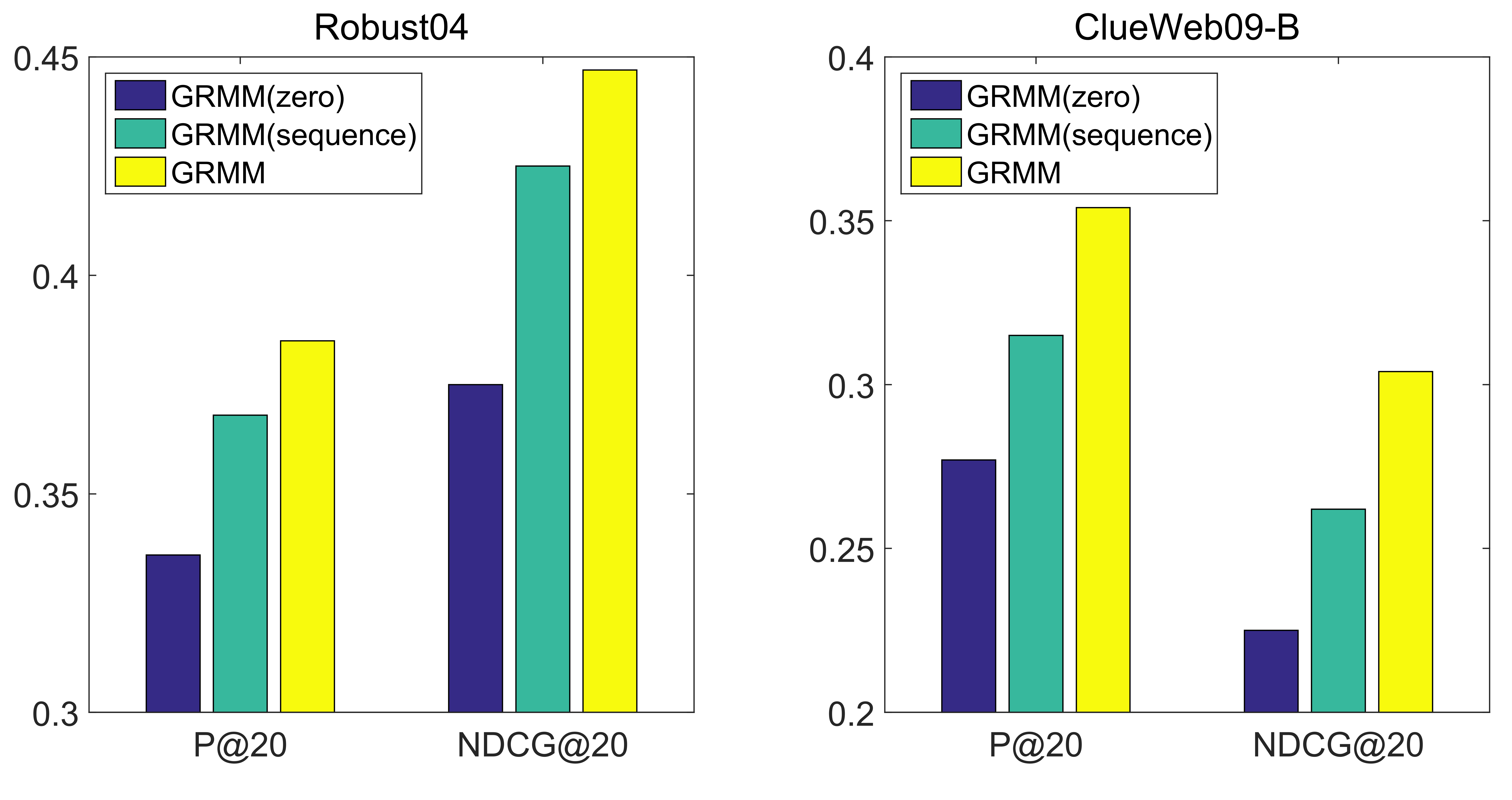}
	\caption{Ablation study on graph structure of GRMM.}
	\label{fig:3} 
\end{figure}

Figure \ref{fig:3} illustrates the comparison between the original GRMM and the alternatives. We can see that:
\begin{itemize}
    \item GRMM (zero matrix) performs inferior to others in all cases. Since it merely depends on the junior term similarities, the model becomes approximate to term-based matching. Without contextualised refinement, some words and their synonyms can be misleading, which makes it even hard to discriminate the actual matching signals. 
    \item GRMM (word sequence) promotes GRMM (zero matrix) by fusing local neighbourhood information but still underperforms the original GRMM by a margin of 2-3 points. This observation resembles some results in Table \ref{tab:2}. It shows that such text format could advantage local context understanding but is insufficient in more comprehensive relationships. 
    \item  From an overall view of the comparison, the document-level word relationships along the graph structure is proved effective for ad-hoc retrieval. Moreover, a relatively greater gain on ClueWeb09-B indicates that longer texts can benefit more from the document-level respective field.
\end{itemize}

\subsection{Study of Neighbourhood Aggregation (RQ2 \& RQ3)}
\label{sec:neighbouraggre}
Figure \ref{fig:4} summarises the experimental performance w.r.t a different number of graph layers. The idea is to investigate the effect of high-order neighbourhood aggregations. For convenience, we notate GRMM-0 for the model with no graph layer, GRMM-1 for the model with a single graph layer, and so forth for the others. From the figure, we find that:

\begin{figure}[h]
	\centering
	\includegraphics[width=.47\textwidth]{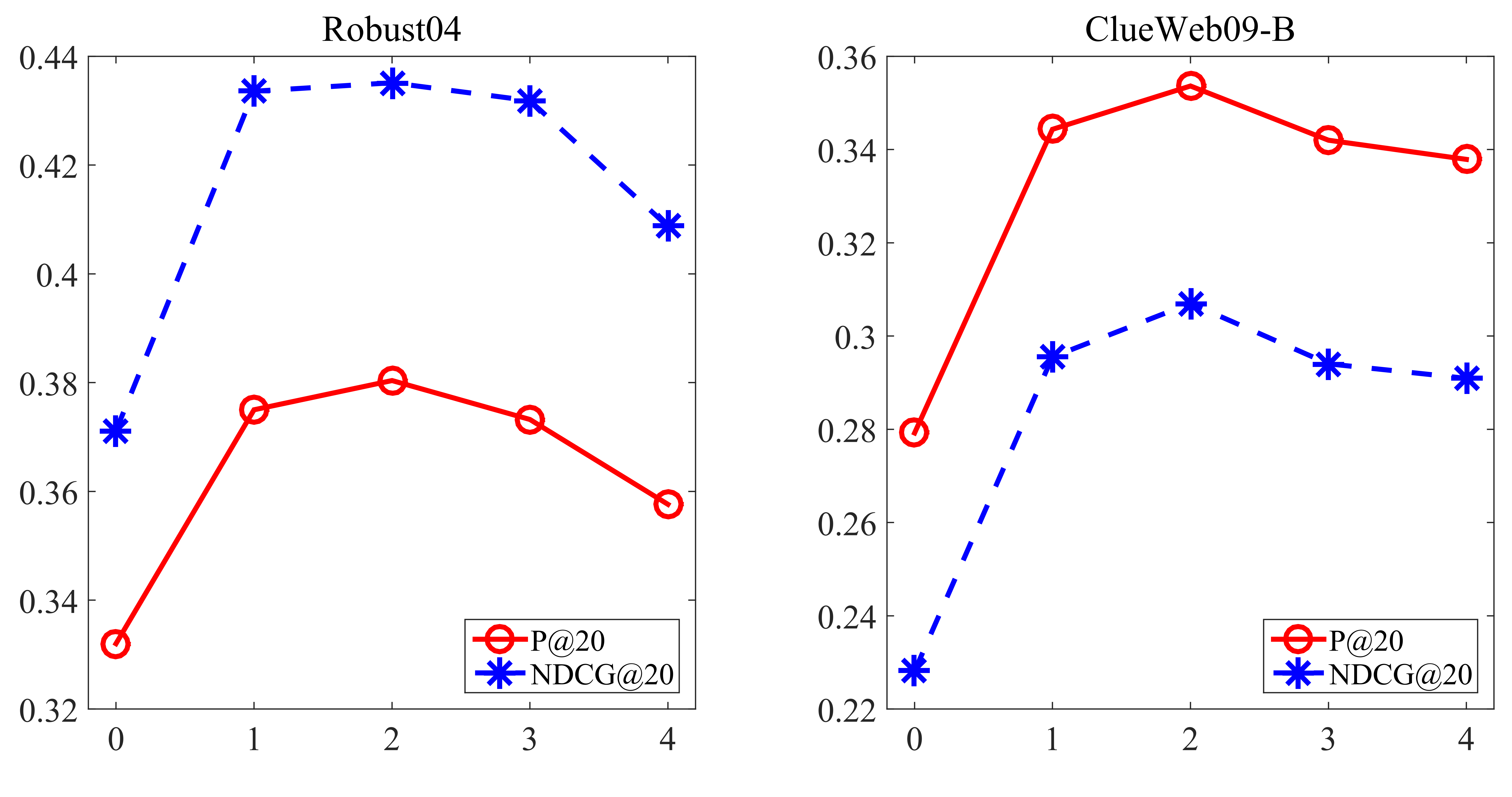}
	\caption{Influence of different graph layer numbers.}
	\label{fig:4} 
\end{figure}

\begin{itemize}
	\item GRMM-1 dramatically boosts the performance against GRMM-0. This observation is consistent with Section \ref{sec:graphstructure} that propagating the information within the graph helps to understand both query-term interaction and document-level word relationships. The exact/similar query-document matching signals are likely to be strengthened or weakened according to intra-document word relationships. 
	\item GRMM-2 improves, not as much though, GRMM-1 by incorporating second-order neighbours. It suggests that the information from 2-hops away also contributes to the term relations. The nodes serving as a bridge can exchange the message from two ends in this way.
	\item However, when further stacking more layers, GRMM-3 and GRMM-4 suffer from slight performance degradation. The reason could be nodes receive more noises from high-order neighbours which burdens the training of parameters. Too much propagation may also lead to the issue of over-smooth \cite{kipf2017semi}. A two-layer propagation seems to be sufficient for capturing useful word relationships.
	\item Overall, there is a tremendous gap between using and not using the contextual information, and the model peaks at layer $t$ = 2 on both datasets. The tendency supports our hypothesis that it is essential to consider term-level interaction and document-level word relationships jointly for ad-hoc retrieval. 
\end{itemize}

\subsection{Study of Graph Readout (RQ3)}
\label{sec:featureelect}
\begin{figure}[h]
	\centering
	\includegraphics[width=.47\textwidth]{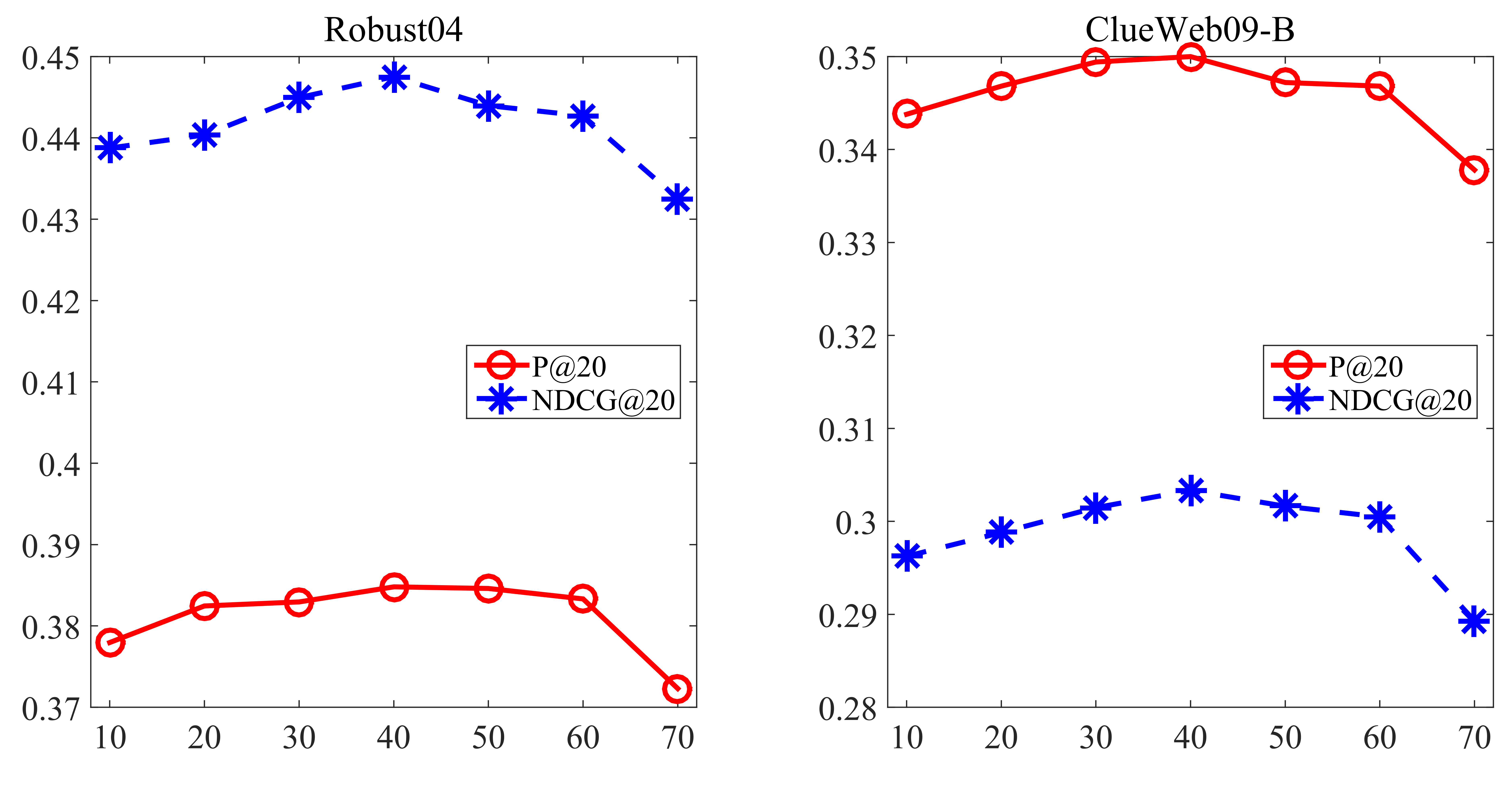}
	\caption{Influence of different $k$ values of $k$-max pooling.}
	\label{fig:5} 
\end{figure}

We also explored the effect of graph readout for each query term. Figure \ref{fig:5} summarises the experimental performance w.r.t different $k$ values of $k$-max-pooling. From the figure, we find that: 
\begin{itemize}
	\item The performance steadily grows from $k$ = 10 to $k$ = 40, which implies that a small feature dimension may limit the representation of terms. By enlarging the $k$ value, the relevant term with more matching signals can distinguish from the irrelevant one with less. 
	\item The trend, however, declines until $k$ = 70, which implies that a large feature dimension may bring negative influence. It can be explained that a large $k$ value may have a bias to the document length, where longer documents tend to have more matching signals. 
	\item Overall, there are no apparent sharp rises and falls in the figure, which tells that GRMM is not that sensitive to the selection of $k$ value. Notably, almost all performances (except $k$ = 70) exceed the baselines in Table \ref{tab:2}, suggesting that determinative matching signals are acquired during the graph-based interactions before feeding into the readout layer. 
\end{itemize}

\section{Conclusion}
In this paper, we introduced a new ad-hoc retrieval approach GRMM which explicitly incorporates document-level word relationships into the matching function. The flexible graph structure allows the model to find more comprehensive matching patterns and less noises. GRMM exceedingly advances the performance over various baselines, where it empirically witnesses an increment by a large margin on longer documents. Further studies exhibited the rationality and effectiveness of GRMM. There are also possible extensions, such as training with large click logs \cite{jiang2016learning} and query descriptions. Another interesting future work is to extend the current graph with lexical or knowledge graphs which might contain more useful information. 

\section*{Acknowledgements}
This work is supported by National Key Research and Development Program (2018YFB1402605, 2018YFB1402600), National Natural Science Foundation of China (U19B2038, 61772528), and Beijing National Natural Science Foundation (4182066).

\bibliography{ref.bib}
\end{document}